\documentclass[conference]{IEEEtran}
\IEEEoverridecommandlockouts

\usepackage{cite}
\usepackage{amsmath,amssymb,amsfonts}
\usepackage{algorithmic}
\usepackage{algorithm}
\usepackage{graphicx}
\usepackage{textcomp}
\usepackage{xcolor}
\def\BibTeX{{\rm B\kern-.05em{\sc i\kern-.025em b}\kern-.08em
    T\kern-.1667em\lower.7ex\hbox{E}\kern-.125emX}}

\usepackage{multirow}
\usepackage{booktabs,makecell,tabularx}
\usepackage{xurl}
\usepackage[caption=false]{subfig}
\usepackage{enumitem}
\usepackage{amsthm}

\definecolor{greyshade}{RGB}{240,240,250}
\definecolor{codeshade}{RGB}{245,245,250}

\UseRawInputEncoding

\definecolor{nodeblue}{RGB}{10,90,180}
\definecolor{commentgreen}{RGB}{2,162,10}
\definecolor{gtmgreen}{RGB}{2,102,10}
\definecolor{eminence}{RGB}{200,95,30}
\definecolor{gclockcyan}{RGB}{50,120,210}
\usepackage{listings}
\usepackage{fancyvrb}
\usepackage{fvextra}
\usepackage{pdfpages}

\begin{document}
\thispagestyle{plain}
\pagestyle{plain}

\title{GaussDB-Global: A Geographically Distributed Database System}
 
\author{\IEEEauthorblockN{Puya Memarzia\IEEEauthorrefmark{1}, 
Huaxin Zhang\IEEEauthorrefmark{1}, Kelvin Ho\IEEEauthorrefmark{1}, 
Ronen Grosman\IEEEauthorrefmark{1}, Jiang Wang\IEEEauthorrefmark{2}}
\IEEEauthorblockA{\IEEEauthorrefmark{1}Huawei Technologies Ltd. Canada, \IEEEauthorrefmark{2}Huawei Technologies Ltd. China\\
Email: \{\IEEEauthorrefmark{1}puya.memarzia1, 
\IEEEauthorrefmark{1}huaxin.zhang,
\IEEEauthorrefmark{1}kelvin.ho,
\IEEEauthorrefmark{1}ronen.grosman,
\IEEEauthorrefmark{2}wangjiang16\}@huawei.com}}

\maketitle
\IEEEpubid{\begin{minipage}{\textwidth}\ \\[40pt]
\footnotesize 10.1109/ICDE60146.2024.00383 \copyright 2024 IEEE. Personal use of this material is permitted. Permission from IEEE must be obtained for all other uses, in any current or future media, including reprinting/republishing this material for advertising or promotional purposes, creating new collective works, for resale or redistribution to servers or lists, or reuse of any copyrighted component of this work in other works.
\end{minipage}}

\begin{abstract}
Geographically distributed database systems use remote replication to protect against regional failures. These systems are sensitive to severe latency penalties caused by centralized transaction management, remote access to sharded data, and log shipping over long distances. To tackle these issues, we present GaussDB-Global, a sharded geographically distributed database system with asynchronous replication, for OLTP applications. To tackle the transaction management bottleneck, we take a decentralized approach using synchronized clocks. Our system can seamlessly transition between centralized and decentralized transaction management, providing efficient fault tolerance and streamlining deployment. To alleviate the remote read and log shipping issues, we support reads on asynchronous replicas with strong consistency, tunable freshness guarantees, and dynamic load balancing. Our experimental results on a geographically distributed cluster show that our approach provides up to 14$\times$ higher read throughput, and 50\% more TPC-C throughput compared to our baseline.
\end{abstract}


\section{Introduction}
\setcounter{page}{1}

Modern database systems are increasingly embracing geographically distributed architectures to support vast numbers of users across multiple regions. Geo-distributed systems rely on remote replication to support strong data availability and minimize data loss in the event of regional disasters. These systems are expected to provide services with high performance and minimal downtime.

\begin{figure}[tbp]
\subfloat[TPC-C Performance Degradation\label{fig:introperfdegrade}]{\includegraphics[width=0.466\linewidth] {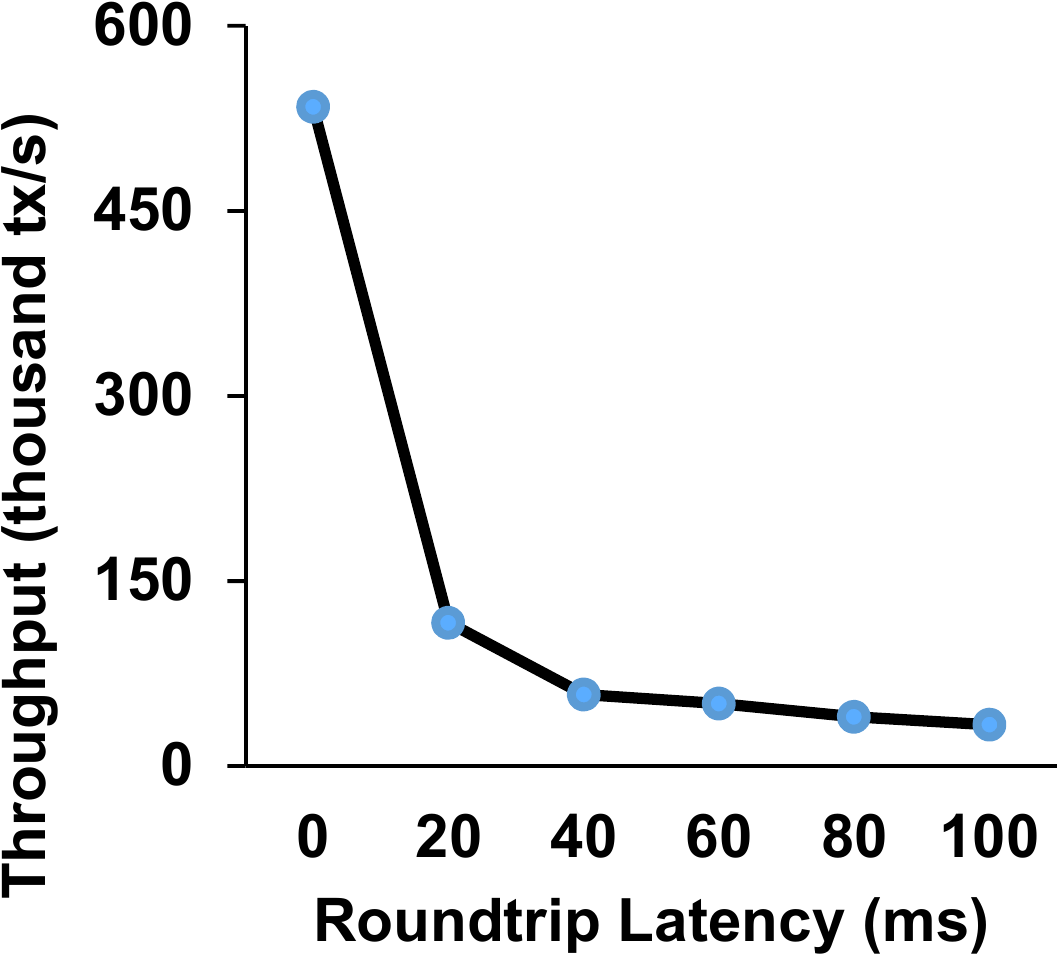}}
\hfill
\subfloat[Global Time and Log Shipping\label{fig:readonreplicaoverview}]{\includegraphics[width=0.466\linewidth]{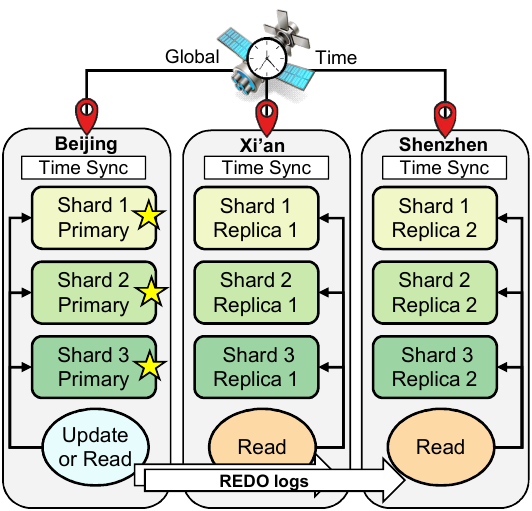}}
\vspace{-1.0mm}
\caption{Geo-distributed Database Overview}
\vspace{-6.0mm}
\label{fig:ror_overview}
\end{figure}

Minimizing cross-region communication is the key to a fast geo-distributed system. Fig.~\ref{fig:introperfdegrade} shows how Online Transaction Processing (OLTP) performance degrades as the system spans across more distant regions. Although decentralized mechanisms such as TrueTime~\cite{spanner2012} can alleviate timestamp assignment overhead, the cost of shipping redo logs (or deltas) across regions can be significant. Fig.~\ref{fig:readonreplicaoverview} illustrates a real-world example with logs shipped between distant cities. In systems that use synchronous replication, write transactions wait on a quorum of replicas before they commit. If the quorum contains remote replicas, transactions must wait longer before they commit, causing performance degradation. Asynchronous replication avoids this issue by not waiting for data to reach distant replicas.

Read-only queries are typically routed to nearby replicas to reduce latency and improve performance. However, asynchronous replication poses new challenges because replicas may contain incomplete or inconsistent data. For sharded databases, each remote shard may have a different amount of redo logs available thus reading the latest data on each replica can produce incorrect results. Customers may tolerate slightly stale data, but will never accept inconsistent data. There has been early research on this problem~\cite{ROS1} but obtaining the freshest possible data while maintaining adequate performance and scalability is challenging. We tackle this issue with a global replica consistency point which guarantees correct reads on asynchronous replicas and can be computed quickly.

Communication overhead from a centralized transaction timestamp server can be avoided by using a decentralized, clock-based solution \cite{spanner2012, cockroachdb2020, sundial, farm2, Yugabyte}. Similarly, our system utilizes a high precision global clock mechanism to support transactions with external serializability. However, seamlessly migrating existing systems to decentralized transaction management is challenging due to fundamental differences in timestamp generation. Furthermore, strong dependence on time synchronization has implications for fault tolerance. Systems that use GPS clocks employ high redundancy~\cite{spanner2012}. In contrast, systems that use commodity hardware clocks pause the cluster if the clock-based mechanism fails~\cite{Yugabyte, cockroachdb2020}. These issues motivated us to develop a novel zero-downtime bi-directional transition mechanism that streamlines deployment and maintenance, and strengthens fault tolerance.

In this paper, we present Huawei's GaussDB-Global (henceforth GlobalDB) as a novel solution for high performance at the geographic scale. We explain how GlobalDB achieves this using a decentralized transaction management system and describe its novel zero downtime transition between centralized and clock-based mechanism. We describe how GlobalDB improves performance further with reads on asynchronous local replicas using a global snapshot of the database. Lastly, we demonstrate with experiments that GlobalDB speeds up geo-distributed workloads with no impact to existing workloads.

Our key contributions are as follows:
\begin{itemize}
\item A decentralized transaction management system based on synchronized global clocks with a mechanism to seamlessly transition between centralized and decentralized transaction management.
\item A flexible asynchronous replication scheme for sharded data that supports reads with guaranteed consistency, bounded freshness, and dynamic load balancing.
\end{itemize}

\section{Background and Related Work}
\label{Sec_Background}

Our system is built on top of Huawei GaussDB~\cite{HuaweiGauss}. In this section we provide some background knowledge on GaussDB, elaborate on the key mechanisms needed to build a geo-distributed system, and compare similar systems.

\subsection{GaussDB Architecture}
\label{Background_GaussArch}

GaussDB is a distributed shared-nothing database system.
A GaussDB cluster consists of computing nodes (CNs), data nodes (DNs), and a lightweight centralized transaction management system called Global Transaction Manager (GTM) which can scale out to over a thousand servers. The CN services client applications, parses queries, generates plans, and coordinates query execution on the DNs. DNs host portions of tables based on the distribution key's hash value or range. Replica DNs are typically placed at remote nodes for high availability. CNs are stateless and hence do not need replicas. The GTM provides timestamps for transaction invocation and commit. These timestamps are used for visibility checking through multi-version concurrency control (MVCC). 

Primary data nodes continuously transmit updates to replica nodes in the form of Redo logs. A transaction may commit once its updates have been propagated to a quorum of replica nodes. If the quorum contains remote replicas, then the database can survive a site-level disaster. Alternatively, a transaction may commit earlier once its updates are replicated to some replicas in the same city. This provides some redundancy but does not protect against a regional disaster.

Although classic shared-nothing systems satisfy the business requirements for single region deployment, geo-distributed deployments pose new challenges. Data is moved or replicated to distant nodes to protect against regional disasters, provide better service to regional clients, and support businesses across geographic regions. Queries may involve expensive inter-node coordination, updates take longer to propagate to replicas, and nodes that are remote from the centralized transaction manager incur a much higher latency when fetching timestamps. We explore some of the related work on these issues.

\subsection{Database Replication}
\label{subsec_relatedwork_replication}

Early research on replicated database reads dates back to the 1990s~\cite{ROS1, ROS2, ROS3}. These approaches have difficulty scaling up due to synchronization overhead \cite{ROS1, ROS2}, and transaction ordering and batching overhead \cite{ROS3}. Compared to those approaches, GlobalDB has a negligible performance impact on the primary data node, does not require a centralized log dispenser, and does not require fine-grained locking when applying Redo logs. Additionally, our system applies Redo logs in parallel which significantly improves log replay speed.

Replication schemes can be divided into two main categories based on how the logs are replicated: synchronous (also known as eager), and asynchronous (also known as lazy or optimistic). In systems with synchronous log replication, transactions wait until all or a quorum of nodes have persisted the update logs to disk~\cite{spanner2012, cockroachdb2020, TiDB2020, Yugabyte}. Synchronous log replication provides strong consistency at the cost of significantly higher update latency. Asynchronous log replication avoids waiting for replica nodes at the cost of weak/eventual data consistency and/or freshness, and a higher risk of data loss~\cite{hbase2011, DynamoDB2022, ConfluxDB2014}. Some database systems can be configured to use either synchronous or asynchronous replication~\cite{HuaweiGauss, MySQLReplication, OceanBase2022}. A third category of systems use epoch-based protocols~\cite{COCO, Geogauss2023}. These systems group transactions within a small time window (the epoch) and defer synchronization with replicas until the epoch boundary. This reduces synchronization overhead compared to synchronous replication. However, aborts and long-running transactions penalize other transactions in the same epoch.

Compared to these systems, GlobalDB is the only sharded geo-distributed relational database with asynchronous physical replication, guaranteed consistency, adjustable freshness, and no negative impact on existing workloads.

\subsection{Distributed Transaction Management}
\label{subsec_relatedwork_gclock}

Database systems with centralized transaction management \cite{HuaweiGauss, OceanBase2022, TiDB2020, DB2, MariaDB} are unsuitable for geo-distributed deployment due to high latency from cross-region communication and unavailability during regional failures. In light of this, modern systems use clock-based approaches to eliminate the overhead of a centralized transaction management system. Spanner~\cite{spanner2012} generates global timestamps using tightly synchronized satellite-connected GPS and atomic clocks with redundancies to provide high availability. GlobalDB employs the same approach for timestamp generation, and introduces a bi-directional transition protocol. In contrast, CockroachDB~\cite{cockroachdb2020} and Yugabyte~\cite{Yugabyte} synchronize clocks without the need for specialized hardware by using software services such as Network Time Protocol (NTP). These systems produce timestamps that are strictly monotonic using a Hybrid Logical Clock (HLC)~\cite{HybridClock2014} combining physical and logical (Lamport) time. Primary nodes append a Lamport timestamp to the commit Redo log indicating the maximum known timestamp of every other shard. Each replica checks if other replicas have applied up to the Lamport timestamp. This approach increases Redo log overhead but saves on deployment cost. Conversely, FaRMv2~\cite{farm2} uses commodity hardware with local time that is synchronized at the data center level.

\section{Global Clock in GaussDB}
\label{Section_GClock}

GaussDB's centralized transaction management can scale up to one thousand servers within a local cluster. However, high-volume timestamp traffic can still negatively impact other types of data flow such as two-phase commits and primary-to-replica Redo log shipping, and remote network traffic degrades performance even further.

We tackle this issue by implementing a global-clock-based (henceforth GClock) algorithm for transaction management. The GClock algorithm is fundamentally the same as Spanner~\cite{spanner2012}. Both meet the following visibility requirements.
\vspace{+1.0mm}
\begin{enumerate}[label=\textbf{R.\arabic*}]
\item If trx$_2$ started after trx$_1$ committed with respect to global time, then trx$_2$ sees trx$_1$'s updates. \label{visreq:1}
\item If trx$_1$ has not committed before trx$_2$ started with respect to global time, trx$_2$ does not see trx$_1$'s updates.\label{visreq:2}
\end{enumerate}
\vspace{+1.0mm}

A transaction gets its GClock timestamp from its computing node's internal clock. The clocks need to be perfectly aligned with each other to satisfy visibility requirements. However, even synchronized clocks can drift apart over time. To resolve this issue, we deploy an accurate and reliable global time source device at each regional cluster. This device includes a GPS receiver and an atomic-clock and is capable of reporting time accurate to within nanoseconds of real time. Machines in the cluster synchronize their clocks with this local global time device every 1 millisecond. Clock deviation is low because synchronization is achieved within 60 microseconds as a TCP round trip, and the CPU's clock drift is bounded within 200 Parts Per Million (PPM)~\cite{farm2}.

A GClock timestamp TS$_{GClock}$ consists of clock time T$_{clock}$ and an error bound T$_{err}$ obtained from the clock synchronization network roundtrip T$_{sync}$ and clock drift T$_{drift}$. The upper and lower bound time is thus obtained from T$_{clock}$ $\pm$ T$_{err}$.

\vspace{-3.5mm}
\begin{align}
\label{eqn-gclock}
    \colorbox{greyshade}{TS$_{GClock}$ = T$_{clock}$ + T$_{err}$\strut}\hspace{+4.00mm}
    \colorbox{greyshade}{T$_{err}$ = T$_{sync}$ + T$_{drift}$\strut}
\end{align}
\vspace{-3.5mm}

Transaction invocations and commits use the following protocol to obtain timestamps.
\vspace{+1.50mm}
\begin{description}
\item[\textbf{Invocation:}]Wait until T$_{clock}$ $>$ TS$_{GClock}$ and begin transaction. Single shard queries bypass this wait by using the node's last committed transaction timestamp.
\item[\textbf{Commit:}]Wait until T$_{clock}$ $>$ TS$_{GClock}$ and commit.
\end{description}
\vspace{+1.50mm}

Following this timestamp protocol, GlobalDB meets the visibility requirements outlined in \ref{visreq:1} and \ref{visreq:2}, thus guaranteeing that all transactions are externally serializable.

GClock improves performance by reducing network and transaction management overhead. Additionally, GClock simplifies deployment at the geographic scale. Compared to prior clock-based transaction management systems, GClock provides a flexible transaction management system which supports both centralized and clock-based modes, and allows online transitioning between these modes without downtime.

\subsection{Migration between GTM and GClock Modes}
\label{subsec-migration}

GTM and GClock transactions are inherently incompatible with each other due to different approaches to timestamp generation. GTM timestamps initially start from zero and increment by one per transaction.

\vspace{-3.5mm}
\begin{align}
\label{eqn-gtm}
    \colorbox{greyshade}{TS$_{GTM}$ = TS$_{GTM}$ + 1\strut}
\end{align}
\vspace{-3.5mm}

In contrast, GClock timestamps use the current epoch time (currently a 10 digit number) which continuously increases even in the absence of new transactions. Even if we initialize GTM to the current time, it is possible for a new GTM transaction to have a smaller timestamp than an older GClock transaction due to the relatively slower growth of GTM timestamps. This presents a problem for migration because correctness requires all transaction timestamps to be monotonically increasing relative to their order.

The most straightforward way to tackle this issue is to block the system from accepting any new transactions and wait until all existing GTM-based transactions have finished and the global time has moved past the last timestamp assigned by the GTM Server. However, this solution entails significant system downtime which is unappealing to customers. Therefore, we propose an online migration approach that allows the co-existence of transactions using different timestamp generation methods. Our approach mitigates the anomalies that may arise from incompatible timestamps.
 
We address this issue in two scenarios: GClock-to-GTM transitions and GTM-to-GClock transitions. A GClock-to-GTM anomaly may occur when a new GTM transaction gets a smaller timestamp than a previously committed GClock transaction. Similarly, a GTM-to-GClock anomaly may occur when clock skew causes a new GClock timestamp to be smaller than a previously committed GTM timestamp.

We resolve these issues with a \textit{DUAL} mode which acts as a bridge between GTM and GClock transactions and can co-exist with both. A DUAL mode timestamp TS$_{DUAL}$ is guaranteed to be larger than both the most recent GTM timestamp TS$_{GTM}$ and clock upper bound. During a transition, new transactions use DUAL mode as an intermediate step to avoid anomalies and satisfy visibility requirements. DUAL mode keeps the system online throughout the transition, continuously accepts new transactions, and maintains correctness.

\vspace{-3.5mm}
\begin{align}
\label{eqn-dual}
    \colorbox{greyshade}{TS$_{DUAL}$ = $\max($TS$_{GTM},$ TS$_{GClock}$) + 1\strut}
\end{align}
\vspace{-3.5mm}

\begin{figure}[tbp]
  \centering
  \includegraphics[width=0.95\linewidth]{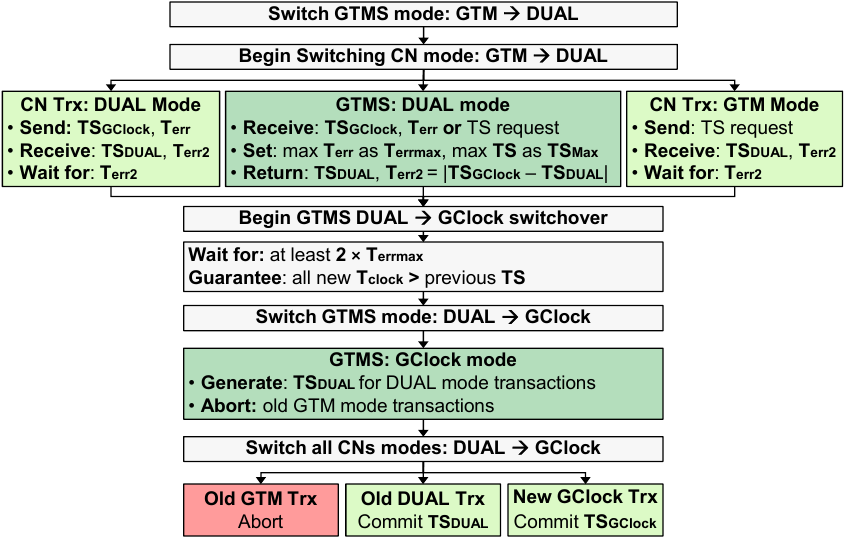}
  \vspace{-2.0mm}
  \caption{GTM to GClock Transition using DUAL mode}
  \label{fig:gtmtogclock}
  \vspace{-4.50mm}
\end{figure}

The protocol to switch the cluster from GTM to GClock using DUAL mode is summarized in Fig~\ref{fig:gtmtogclock}. We begin by switching the GTM server (GTMS) and then each CN to DUAL mode. During this transition, the GTM server will service timestamp requests for both GTM and DUAL mode CNs. The GTM server also tracks the maximum issued timestamp and error bound until all nodes acknowledge the switch to DUAL mode. DUAL mode transactions first obtain a GClock timestamp and then communicate with the GTM server to receive a commit timestamp and a wait duration that avoids anomalies with existing GTM transactions. The GTM must remain in DUAL mode for $2\times$ the maximum error bound observed during the GTM to DUAL mode transition period. Only after this wait time can the cluster begin transitioning from DUAL mode to GClock mode, again starting from the GTM server followed by the CNs. All new transactions from this point on will use GClock mode. The wait time ensures that new GClock timestamps will be larger than all previous timestamps. All running DUAL mode transactions can commit safely. Old GTM mode transactions that try to commit after the cluster has  transitioned to GClock mode will abort.

During the transition, the GTM server is in DUAL mode but it is possible to have a combination of GTM, DUAL, and GClock transactions running concurrently because the CNs might not switch modes at precisely the same moment. A wait time is needed during DUAL mode to ensure correctness, otherwise the GTM-to-GClock transitions may cause visibility issues. Consider the following example.

\vspace{-3mm}
\begin{lstlisting}[
    backgroundcolor = \color{codeshade},
    basicstyle=\small, %or \small or \footnotesize etc.
    caption={Example showing why GTM transactions wait in DUAL mode},
]
GTMS   Running in DUAL mode
Node1  Running Trx1 in GTM mode
Node2  GTM mode = DUAL mode
Node3  GTM mode = DUAL mode
Node1  GTM mode = DUAL mode
Node2  DUAL mode = GClock mode
Node3  Send large GClock timestamp ts3 to GTMS
GTMS   Raise internal timestamp to ts3
Node1  Trx1 gets large DUAL mode timestamp ts1 > ts3
       from GTMS and Commit Trx1 without waiting
Node2  Trx2 starts with timestamp ts2 < ts1
       Trx2 cannot see Trx1's committed update
\end{lstlisting}
\vspace{-1mm}

To avoid this issue, GTM mode transactions must wait at commit if the GTM server is in DUAL mode. As shown in Fig~\ref{fig:gtmtogclock}, this wait time is double the largest error bound received by the GTM server during the transition.

\begin{figure}[tbp]
  \centering
  \includegraphics[width=0.97\linewidth]{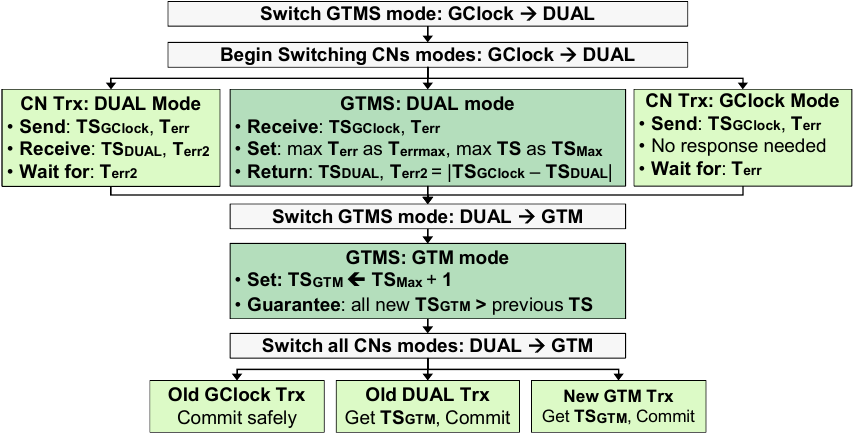}
  \vspace{-2.0mm}
  \caption{GClock to GTM Transition using DUAL mode}
  \label{fig:gclocktogtm}
  \vspace{-4.50mm}
\end{figure}

Transitioning from GClock back to GTM mode is illustrated in Fig~\ref{fig:gclocktogtm}. This scenario may occur if there is a clock or synchronization issue. The system can safely switch to GTM mode and  shift back to GClock mode once the issue is resolved. The logic for GClock-to-GTM transition is similar albeit slightly simpler than a GTM-to-GClock transition. The GTM server keeps track of the largest GClock timestamp issued so far. As a result, no old transactions will need to abort because the GTM server will issue timestamps that are larger than the largest GClock timestamp issued until the moment of transition. This also eliminates the need to wait while in DUAL mode, and nodes can begin switching to GTM mode as soon as all nodes have switched to DUAL mode.

\section{Reads on Replica}
\label{Section_ROR}
 
In GlobalDB, replica nodes can be used to fulfill read-only queries. This greatly reduces network latency if the client or the computing node is physically closer to the replica node than the primary node. If a primary node fails, its replica nodes can continue to serve read-only queries until the failed primary node recovers, or a replica node is promoted to replace the primary node. Allowing replica reads also helps to distribute the load from primary nodes to less busy replica nodes.

\subsection{Read with Consistency at Replica}
\label{subsec_ror}
GlobalDB is a shared-nothing architecture where relations may be distributed across multiple shards. Each shard has a primary node that accepts read and write operations and one or more replica nodes that are read-only (Fig.~\ref{fig:readonreplicaoverview}). Redo logs are shipped asynchronously, and the speed of applying Redo logs at each replica nodes may be different. As a result, it is possible for different shards of the same relation to have different levels of freshness at the replicas. Therefore, we need a consistent snapshot of the database, representing data that is available on all replicas at a point of time in history that is as close to the present as possible. We call this point in time the \textit{Replica Consistency Point} (RCP).

Finding and maintaining the RCP is non-trivial and related research dates back to the early 1990s~\cite{ROS2,ROS3}. Our approach involves finding the largest commit timestamp that is available on all replica nodes. Transactions committed before the RCP timestamp are visible, and any partial transactions and transactions with unfulfilled dependencies are invisible.

\begin{figure}[tbp]
  \centering
  \includegraphics[width=0.9\linewidth]{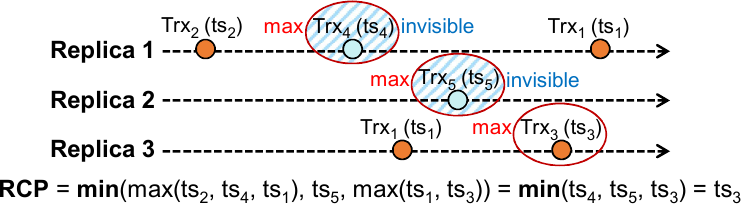}
  \vspace{-2.0mm}
  \caption{Replica Consistency Point Calculation}
  \label{fig:readonreplicaconsistencypoint}
  \vspace{-5.0mm}
\end{figure}

The example in Fig.~\ref{fig:readonreplicaconsistencypoint} illustrates a scenario involving three different replicated shards. Each replica has a stream of incoming Redo logs with transaction commit timestamps ranging from $ts_1$ to $ts_5$ in chronological order. Using the algorithm, GlobalDB picks the maximum commit timestamps $ts_4$ from Replica 1, $ts_5$ from Replica 2, and $ts_3$ from Replica 3, shown circled in the figure. Thus the RCP timestamp is calculated as $min \{ts_4, ts_5, ts_3\} = ts_3$. All ROR queries will return transactions with a lesser or equal timestamp to ts$_3$, meaning Trx$_1$, Trx$_2$ and Trx$_3$ are the only visible transactions at this point. This is accurate because Trx$_4$ may have more than one shard involved, and we do not know if its Redo logs will arrive on Replica 2 or Replica 3. Trx$_5$ may depend on Trx$_4$ as it has a larger timestamp, so it cannot be visible. On Replica 1 Trx$_1$'s commit timestamp is smaller than Trx$_2$'s, meaning it does not depend on Trx$_2$, even if its Redo log appears after Trx$_2$'s redo log. Therefore Trx$_1$, Trx$_2$ and Trx$_3$ are either single shard transactions, or committed, or pending their final commit Redo log as explained below.

Although commit timestamps increase monotonically, the order in which they are written to the Redo log is not necessarily ascending. This is because getting the timestamp from GClock or GTM and writing the timestamp as a commit Redo record may occur out of order due to thread context switching. Therefore, to provide correct visibility, we wait on tuples that are associated with in-progress transactions until they either commit or abort. This safe-guard is implemented by writing a special PENDING\_COMMIT redo log at the primary before the transaction gets its invocation timestamp. This in turn locks the associated tuples on the replica node. Similarly, for two-phase commit transactions, a prepared transaction's visibility check at the replica is blocked until a COMMIT PREPARED or ABORT PREPARED record is replayed.

Each replica keeps track of its maximum commit timestamp. A CN is selected at the remote site to periodically collect the maximum  timestamps from the replicas, calculate the RCP, and distribute it to other CNs. If the CN goes down, a different CN is selected to take over the RCP calculation. This approach has two benefits. First, it prevents the RCP from moving backwards from the perspective of client applications because clients may get routed to different CNs for reasons such as load balancing and failover. Second, it allows CNs to use remote replicas for reads. This is useful if the local replica is down, overloaded, or stale. Reading from a remote replica may still be faster than a remote primary. We describe this node selection algorithm in Section~\ref{Subsec_ROR_node_select}.

Not all transactions involve all data nodes, thus a replica node's maximum timestamp could lag behind when it does not receive any transactions to replay. A \textit{heartbeat} transaction is periodically sent from the CN to all replicas to guarantee that the max commit timestamp always increases. From a client application's point of view, the RCP increases monotonically and consecutive ROR queries always show data with equal or greater freshness than previous queries.

When a Data Definition Language (DDL) statement (such as CREATE TABLE or DROP INDEX) is executed, it is expected to be visible and effective on subsequent queries. Due to the inherent delay in replaying logs to the replica nodes, we take additional measures to ensure the ROR queries are consistent with any relevant DDL statements. 
As such, we allow ROR queries if at least one of the following conditions are true:

\begin{enumerate}
\item The RCP is greater than the largest DDL timestamp. This means all DDLs have been replayed on all replicas. We skip the second check if this one passes.
\item The RCP is greater than the DDL timestamp for each table that is involved in the ROR query.
\end{enumerate}

\subsection{ROR Node Selection}
\label{Subsec_ROR_node_select}
In a geo-distributed cluster, the same data may be available from a multitude of nodes with varying levels of freshness, performance, and health. Reading from any replica located within the same region typically achieves the lowest network latency, but it does not distinguish between different replicas in the same region or consider each node's data freshness, load, and failure state. To solve this issue, we propose a dynamic node selection mechanism that picks the best nodes for each query based on per-node metrics that the CN tracks.

\begin{figure}[tbp]
  \centering
  \includegraphics[width=0.53\linewidth]{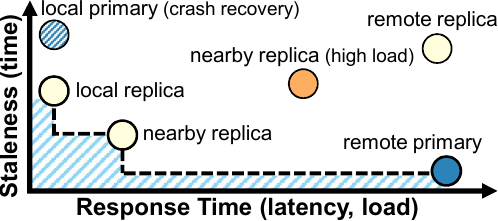}
  \vspace{-2.00mm}
  \caption{ROR dynamic node selection using skyline}
  \label{fig:costbasednodeselection}
  \vspace{-4.50mm}
\end{figure}

GlobalDB automatically detects failed or overloaded nodes and reroutes queries to other nodes. This rerouting is periodically done in the background, allowing GaussDB to achieve load balancing and respond to changes in node status. Fig.~\ref{fig:costbasednodeselection} illustrates our cost-based node selection using how much data a replica node has replayed (data staleness) and how promptly this node responds to queries (latency and load). Each computing node periodically refreshes this metric to form a skyline of candidate nodes. When running under GClock mode, replica staleness is measured by comparing the last committed transaction's timestamp against the current time. When running under GTM mode, we estimate the staleness based on the gap between the RCP and the last committed timestamp, and the rate at which new timestamps were issued during the last interval. Given a query with a bounded staleness requirement, the computing node picks a set of replicas from the skyline to answer the query with minimal latency.

The cost/staleness-based algorithm for picking replica nodes helps us to dynamically balance workload and provide high availability. For example, we may offload a busy primary node's reads to a replica node, or we may swap out a replica node for a different one if its response time goes up. When a node crashes, it is automatically excluded from the skyline.

\begin{figure*}[tbp]
\centering
\includegraphics[scale=0.2]{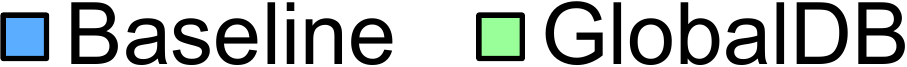}
\vspace{-3.00mm}
\hfill

\subfloat[TPC-C - Synchronous\label{fig:result_3city_baselinevsglobaldb}]{\includegraphics[scale=0.23]{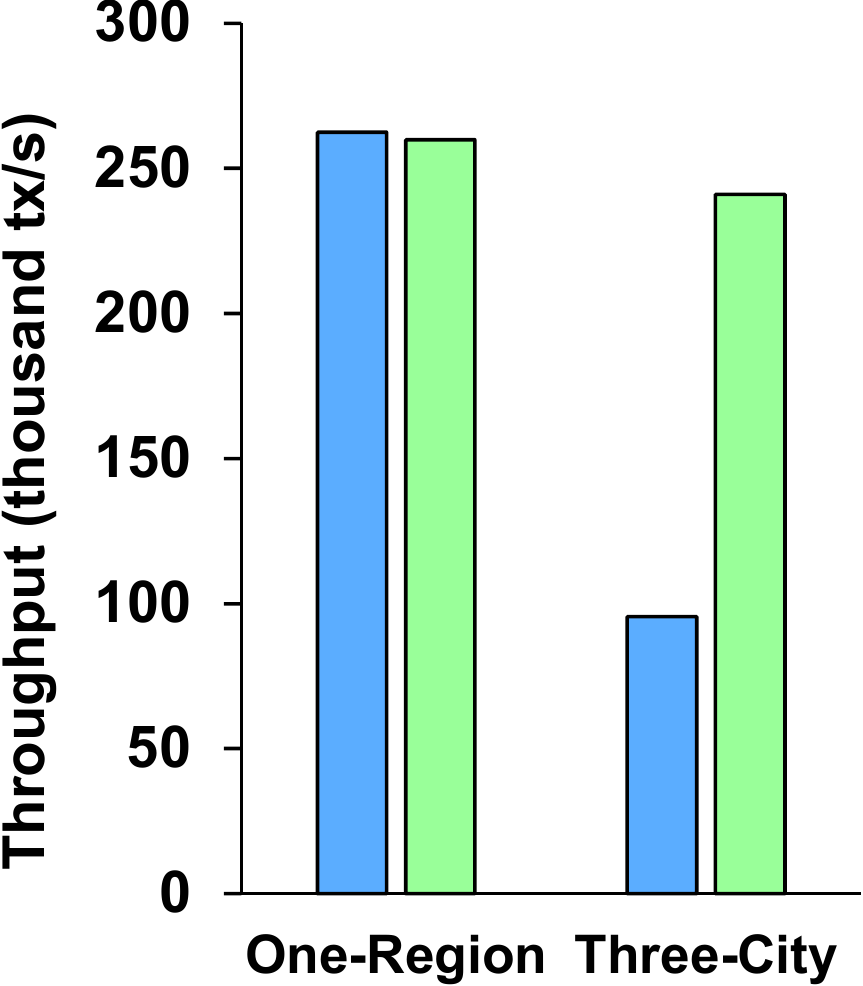}}
\hfill
\subfloat[TPC-C - Asynchronous\label{fig:tpcc_gclockvsgtm_samerack}]{\includegraphics[scale=0.23]{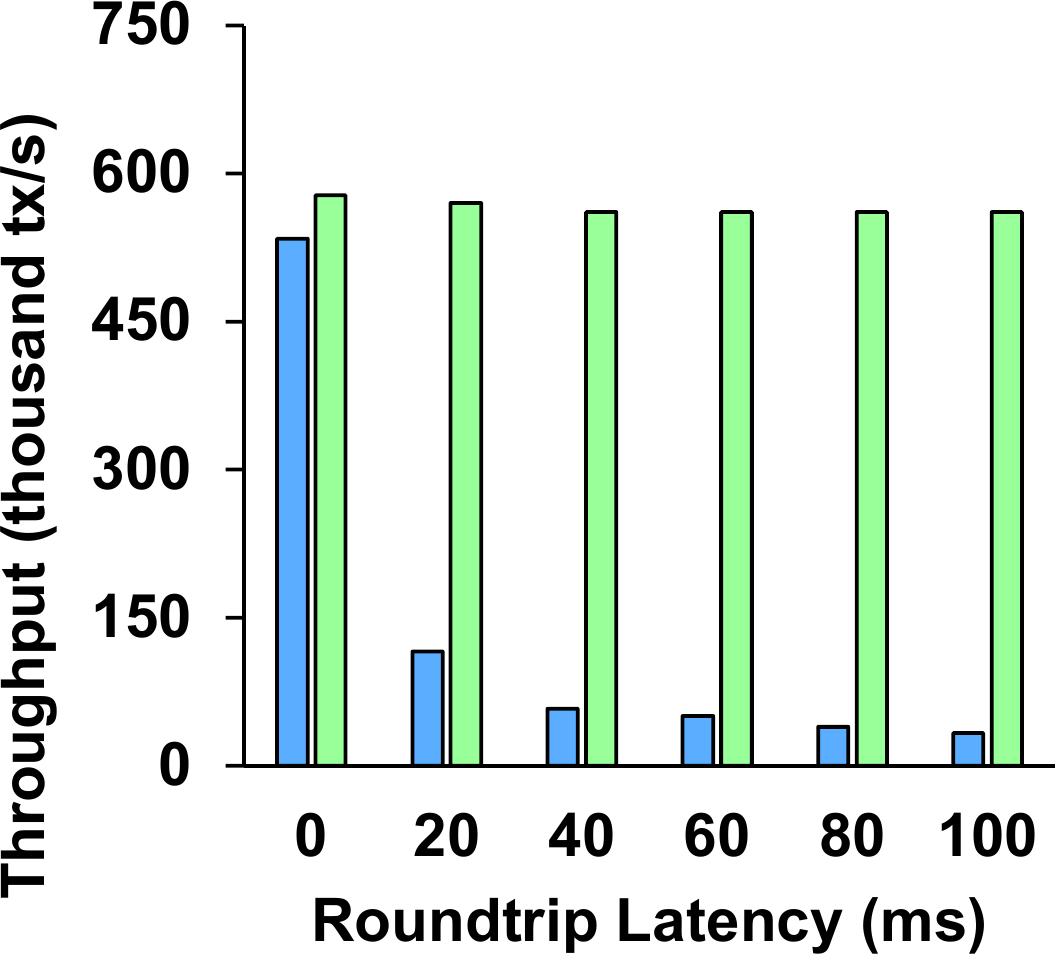}}
\hfill
\subfloat[TPC-C - Modified (Read-only)\label{fig:result_tpcc_readonly}]{\includegraphics[scale=0.23]{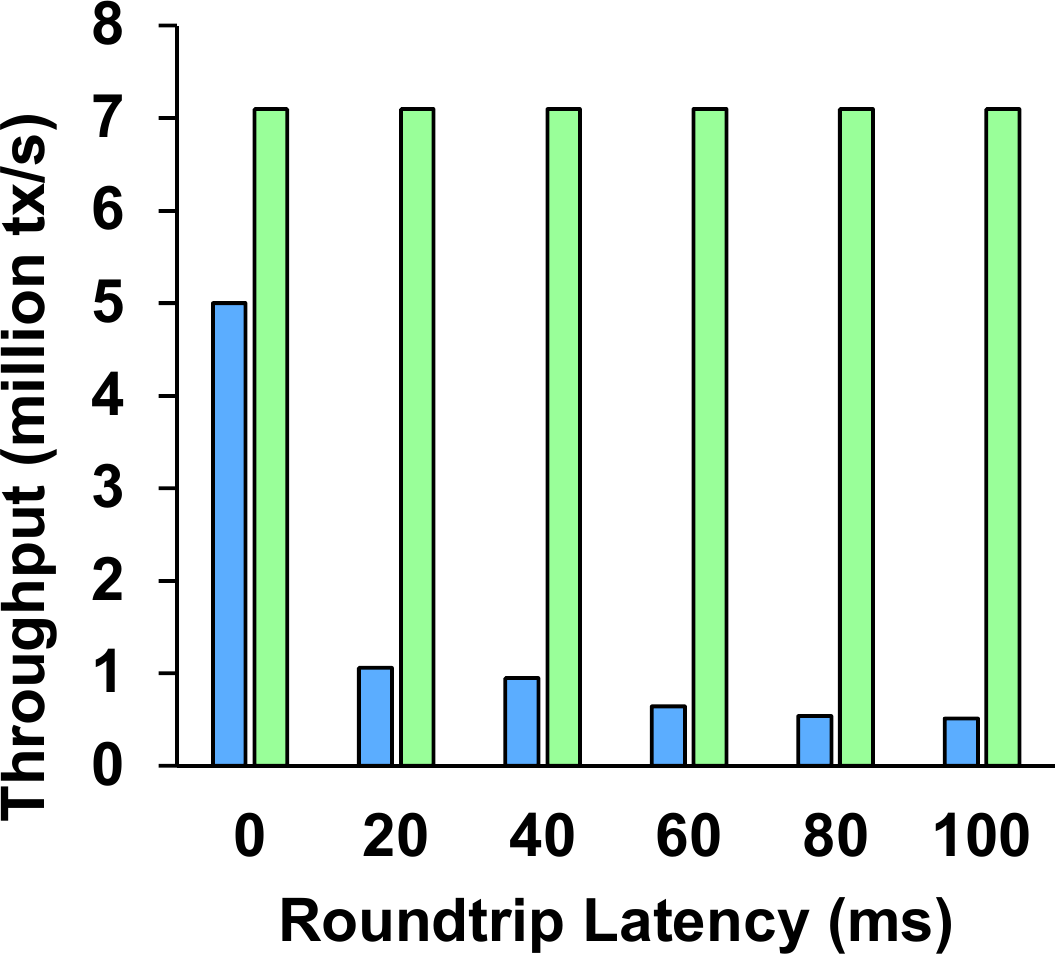}}
\hfill
\subfloat[Sysbench Point Select\label{fig:result_sysbench_pointread}]{\includegraphics[scale=0.23]{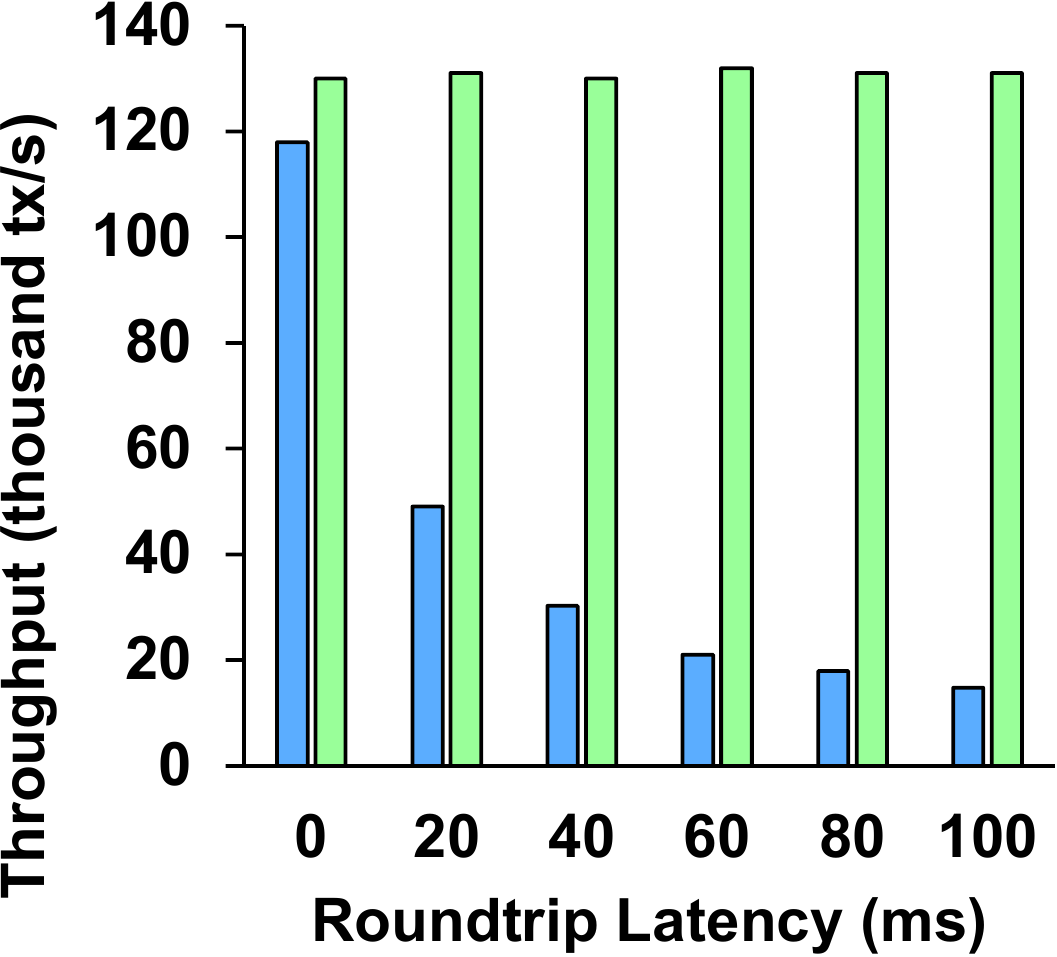}}
\vspace{-1.5mm}
\caption{Experimental results}
\label{fig:ror_results}
\vspace{-4mm}
\end{figure*}

\section{Performance Evaluation}
\label{Sec_Eval}

Our experimental setup consists of a cluster deployed within a single data center with simulated network delay (henceforth One-Region), and a geographically distributed cluster deployed in three different cities (henceforth Three-City).

The One-Region cluster consists of three Huawei ARM64 TaiShan servers all housed within the same server rack and connected via 10 gigabit Ethernet. Each server is equipped with two Huawei Kunpeng 920~\cite{HuaweiTaishan1} CPUs and 256GB of DRAM. The Three-City cluster includes a high precision time device in each region, and consists of three Intel x86 servers equipped with two Intel Xeon E5-2680 v2 CPUs, and 188GB of DRAM. The servers are located in Xi'an, Langzhong, and Dongguan, forming a triangle with 25ms, 35ms, and 55ms latency on each edge. Both clusters run EulerOS~\cite{euleros}.

Each cluster has a total of three computing nodes, six data nodes, and 12 replica nodes. We evaluate performance using TPC-C~\cite{TPC-C} and Sysbench~\cite{Sysbench}. Unless otherwise noted, we run the full TPC-C benchmark with all five transaction types, configured with 600 warehouses and 600 client terminals. The Sysbench experiments are configured with 250 tables with 25000 rows each and 600 client threads.

\subsection{Transaction Management and Log Shipping}
\label{results_gclock}

In this set of experiments we demonstrate GlobalDB's efficient transaction management and redo log shipping.

Real customer workloads have some degree of physical affinity. In light of this, we modify our workloads to control the proportion of remote transactions. To quantify improvement from each feature separately, we start with $100\%$ local transactions to evaluate GClock. In this scenario, the sources of performance degradation are limited to transaction management and log replication. We later increase the percentage of remote transactions in Section~\ref{results_ror} to evaluate ROR.

On the Three-City cluster, the network bandwidth between cities is considerably lower compared to the One-Region cluster. This increases transaction latency because Redo logs are buffered for longer before they can be transmitted. We examine this network overhead by switching GlobalDB to synchronous replication and running TPC-C. We also collocate the GTM server on the machine with the lowest mean latency to the other machines. The results in Fig.~\ref{fig:result_3city_baselinevsglobaldb} show that the baseline's throughput decreases by about two thirds. GlobalDB eliminates transaction management overhead by using GClock, and improves network overhead by compressing the Redo logs with LZ4 compression~\cite{LZ4}, utilizing TCP BBR for more aggressive congestion control \cite{TCPcongestioncontrol}, and disabling Nagel's buffering algorithm to reduce receiver acknowledgement latency \cite{nagle1984congestion}. 
Once we deploy GlobalDB on the Three-City cluster, throughput increases to $91\%$ of the One-Region cluster. Furthermore, GlobalDB does not suffer a performance penalty when directly deployed on the One-Region cluster.

In Figs.~\ref{fig:tpcc_gclockvsgtm_samerack},~\ref{fig:result_tpcc_readonly},~\ref{fig:result_sysbench_pointread} we simulate some network delay on our One-Region cluster using Linux Traffic Control (tc)~\cite{trafficcontrol}. To emphasize the transaction management network overhead, we show the throughput of a node that is not co-located with the GTM server.
In Fig.~\ref{fig:tpcc_gclockvsgtm_samerack} 
we observe that baseline GaussDB's performance degrades by up to $90\%$ when a 100ms network delay is added between the machines. GlobalDB achieves the same throughput regardless of network delay.

\subsection{Read Performance}
\label{results_ror}

Read throughput emphasizes the benefits of GlobalDB's ROR feature. Here we modify TPC-C to only run the Order-status and Stock-level transactions thus turning it into a read-only benchmark. $50\%$ of transactions are configured to be multi-shard. In  Fig.~\ref{fig:result_tpcc_readonly} we see that TPC-C read throughput improves by up to $14\times$ on GlobalDB compared to the baseline due to reading from replicas and reduced transaction management overhead. In the Sysbench Point Select workload, 2/3 of the tuples are fetched from a remote node. As shown in Fig.~\ref{fig:result_sysbench_pointread}, GlobalDB improves Sysbench read throughput by up to $8.9\times$ over the baseline due to reading from local replicas.

\section{Conclusion}
\label{Sec_Conclusion}

GaussDB-Global (GlobalDB) is a sharded, geographically distributed relational database system designed for high transaction throughput, low latency reads, and high availability. Our system achieves high performance using a decentralized clock-based transaction management system, reads from asynchronous replicas, and redo log shipping optimizations. GlobalDB guarantees external consistency using either decentralized or centralized transaction management. Our novel DUAL mode and bi-directional transition mechanism allows geo-distributed features to be activated on a live system without the need to take it offline. It also keeps the system fully operational in the event of a clock synchronization failure. GlobalDB's novel Read-On-Replica feature improves read performance with guaranteed consistency, adjustable freshness, and zero impact to write performance. Different replication schemes and different levels of disaster recovery provide flexibility to meet different customer requirements. Our experimental results show that GlobalDB can achieve performance that approaches a co-located cluster without any performance regressions for existing workloads. For future work, we are considering transparent load balancing based on geographical access patterns, self-assembling geo-distributed clusters to assist with deployment, and synchronous replicated tables that co-exist with asynchronous tables to meet specific business requirements by trading off update performance in favor of maximizing freshness and read performance.

\IEEEtriggeratref{42}

\bibliographystyle{IEEEtran}
\bibliography{IEEEabrv,globaldb}

\begin{thebibliography}{10}
\providecommand{\url}[1]{#1}
\csname url@samestyle\endcsname
\providecommand{\newblock}{\relax}
\providecommand{\bibinfo}[2]{#2}
\providecommand{\BIBentrySTDinterwordspacing}{\spaceskip=0pt\relax}
\providecommand{\BIBentryALTinterwordstretchfactor}{4}
\providecommand{\BIBentryALTinterwordspacing}{\spaceskip=\fontdimen2\font plus
\BIBentryALTinterwordstretchfactor\fontdimen3\font minus \fontdimen4\font\relax}
\providecommand{\BIBforeignlanguage}[2]{{%
\expandafter\ifx\csname l@#1\endcsname\relax
\typeout{** WARNING: IEEEtran.bst: No hyphenation pattern has been}%
\typeout{** loaded for the language `#1'. Using the pattern for}%
\typeout{** the default language instead.}%
\else
\language=\csname l@#1\endcsname
\fi
#2}}
\providecommand{\BIBdecl}{\relax}
\BIBdecl

\bibitem{spanner2012}
\BIBentryALTinterwordspacing
J.~C. Corbett, J.~Dean, M.~Epstein, A.~Fikes, C.~Frost, J.~Furman, S.~Ghemawat, A.~Gubarev, C.~Heiser, P.~Hochschild, W.~Hsieh, S.~Kanthak, E.~Kogan, H.~Li, A.~Lloyd, S.~Melnik, D.~Mwaura, D.~Nagle, S.~Quinlan, R.~Rao, L.~Rolig, Y.~Saito, M.~Szymaniak, C.~Taylor, R.~Wang, and D.~Woodford, ``Spanner: {Google{\textquoteright}s} {Globally-Distributed} database,'' in \emph{10th USENIX Symposium on Operating Systems Design and Implementation (OSDI 12)}.\hskip 1em plus 0.5em minus 0.4em\relax Hollywood, CA: USENIX Association, Oct. 2012, pp. 261--264. [Online]. Available: \url{https://www.usenix.org/conference/osdi12/technical-sessions/presentation/corbett}
\BIBentrySTDinterwordspacing

\bibitem{ROS1}
\BIBentryALTinterwordspacing
C.~A. Polyzois and H.~Garcia-Molina, ``Evaluation of remote backup algorithms for transaction processing systems,'' in \emph{Proceedings of the 1992 ACM SIGMOD International Conference on Management of Data}, ser. SIGMOD '92.\hskip 1em plus 0.5em minus 0.4em\relax New York, NY, USA: Association for Computing Machinery, 1992, p. 246–255. [Online]. Available: \url{https://doi.org/10.1145/130283.130321}
\BIBentrySTDinterwordspacing

\bibitem{cockroachdb2020}
\BIBentryALTinterwordspacing
R.~Taft, I.~Sharif, A.~Matei, N.~VanBenschoten, J.~Lewis, T.~Grieger, K.~Niemi, A.~Woods, A.~Birzin, R.~Poss, P.~Bardea, A.~Ranade, B.~Darnell, B.~Gruneir, J.~Jaffray, L.~Zhang, and P.~Mattis, ``Cockroachdb: The resilient geo-distributed sql database,'' in \emph{Proceedings of the 2020 ACM SIGMOD International Conference on Management of Data}, ser. SIGMOD '20.\hskip 1em plus 0.5em minus 0.4em\relax New York, NY, USA: Association for Computing Machinery, 2020, p. 1493–1509. [Online]. Available: \url{https://doi.org/10.1145/3318464.3386134}
\BIBentrySTDinterwordspacing

\bibitem{sundial}
Y.~Li, G.~Kumar, H.~Hariharan, H.~Wassel, P.~Hochschild, D.~Platt, S.~Sabato, M.~Yu, N.~Dukkipati, P.~Chandra, and A.~Vahdat, ``Sundial: Fault-tolerant clock synchronization for datacenters,'' in \emph{Proceedings of the 14th USENIX Conference on Operating Systems Design and Implementation}, ser. OSDI'20.\hskip 1em plus 0.5em minus 0.4em\relax USA: USENIX Association, 2020.

\bibitem{farm2}
\BIBentryALTinterwordspacing
A.~Shamis, M.~Renzelmann, S.~Novakovic, G.~Chatzopoulos, A.~Dragojevi\'{c}, D.~Narayanan, and M.~Castro, ``Fast general distributed transactions with opacity,'' in \emph{Proceedings of the 2019 International Conference on Management of Data}, ser. SIGMOD '19.\hskip 1em plus 0.5em minus 0.4em\relax New York, NY, USA: Association for Computing Machinery, 2019, p. 433–448. [Online]. Available: \url{https://doi.org/10.1145/3299869.3300069}
\BIBentrySTDinterwordspacing

\bibitem{Yugabyte}
\BIBentryALTinterwordspacing
(2020) Yugabytedb: Distributed sql database. Yugabyte, inc. [Online]. Available: \url{https://www.yugabyte.com/}
\BIBentrySTDinterwordspacing

\bibitem{HuaweiGauss}
\BIBentryALTinterwordspacing
(2023) Gaussdb distributed relational database system. Huawei. [Online]. Available: \url{https://www.huaweicloud.com/intl/en-us/product/gaussdb.html}
\BIBentrySTDinterwordspacing

\bibitem{ROS2}
R.~P. King, N.~Halim, H.~Garcia-Molina, and C.~A. Polyzois, ``Management of a remote backup copy for disaster recovery,'' \emph{ACM Transactions on Database Systems (TODS)}, vol.~16, no.~2, pp. 338--368, 1991.

\bibitem{ROS3}
C.~A. Polyzois and H.~Garcia-Molina, ``Evaluation of remote backup algorithms for transaction-processing systems,'' \emph{ACM Transactions on Database Systems (TODS)}, vol.~19, no.~3, pp. 423--449, 1994.

\bibitem{TiDB2020}
\BIBentryALTinterwordspacing
D.~Huang, Q.~Liu, Q.~Cui, Z.~Fang, X.~Ma, F.~Xu, L.~Shen, L.~Tang, Y.~Zhou, M.~Huang, W.~Wei, C.~Liu, J.~Zhang, J.~Li, X.~Wu, L.~Song, R.~Sun, S.~Yu, L.~Zhao, N.~Cameron, L.~Pei, and X.~Tang, ``Tidb: A raft-based htap database,'' \emph{Proc. VLDB Endow.}, vol.~13, no.~12, p. 3072–3084, aug 2020. [Online]. Available: \url{https://doi.org/10.14778/3415478.3415535}
\BIBentrySTDinterwordspacing

\bibitem{hbase2011}
M.~N. Vora, ``Hadoop-hbase for large-scale data,'' in \emph{Proceedings of 2011 International Conference on Computer Science and Network Technology}, vol.~1, 2011, pp. 601--605.

\bibitem{DynamoDB2022}
\BIBentryALTinterwordspacing
M.~Elhemali, N.~Gallagher, N.~Gordon, J.~Idziorek, R.~Krog, C.~Lazier, E.~Mo, A.~Mritunjai, S.~Perianayagam, T.~Rath, S.~Sivasubramanian, J.~C.~S. III, S.~Sosothikul, D.~Terry, and A.~Vig, ``Amazon {DynamoDB}: A scalable, predictably performant, and fully managed {NoSQL} database service,'' in \emph{2022 USENIX Annual Technical Conference (USENIX ATC 22)}.\hskip 1em plus 0.5em minus 0.4em\relax Carlsbad, CA: USENIX Association, Jul. 2022, pp. 1037--1048. [Online]. Available: \url{https://www.usenix.org/conference/atc22/presentation/elhemali}
\BIBentrySTDinterwordspacing

\bibitem{ConfluxDB2014}
\BIBentryALTinterwordspacing
P.~Chairunnanda, K.~Daudjee, and M.~T. \"{O}zsu, ``Confluxdb: Multi-master replication for partitioned snapshot isolation databases,'' \emph{Proc. VLDB Endow.}, vol.~7, no.~11, p. 947–958, jul 2014. [Online]. Available: \url{https://doi.org/10.14778/2732967.2732970}
\BIBentrySTDinterwordspacing

\bibitem{MySQLReplication}
\BIBentryALTinterwordspacing
(2023) Mysql 8.0 reference manual. Oracle. [Online]. Available: \url{https://dev.mysql.com/doc/refman/8.0/en/replication.html}
\BIBentrySTDinterwordspacing

\bibitem{OceanBase2022}
\BIBentryALTinterwordspacing
Z.~Yang, C.~Yang, F.~Han, M.~Zhuang, B.~Yang, Z.~Yang, X.~Cheng, Y.~Zhao, W.~Shi, H.~Xi, H.~Yu, B.~Liu, Y.~Pan, B.~Yin, J.~Chen, and Q.~Xu, ``Oceanbase: A 707 million tpmc distributed relational database system,'' \emph{Proc. VLDB Endow.}, vol.~15, no.~12, p. 3385–3397, aug 2022. [Online]. Available: \url{https://doi.org/10.14778/3554821.3554830}
\BIBentrySTDinterwordspacing

\bibitem{COCO}
Y.~Lu, X.~Yu, L.~Cao, and S.~Madden, ``Epoch-based commit and replication in distributed oltp databases,'' \emph{Proc. VLDB Endow.}, vol.~14, no.~5, p. 743–756, jan 2021.

\bibitem{Geogauss2023}
W.~Zhou, Q.~Peng, Z.~Zhang, Y.~Zhang, Y.~Ren, S.~Li, G.~Fu, Y.~Cui, Q.~Li, C.~Wu, S.~Han, S.~Wang, G.~Li, and G.~Yu, ``Geogauss: Strongly consistent and light-coordinated oltp for geo-replicated sql database,'' \emph{Proc. ACM Manag. Data}, vol.~1, no.~1, may 2023.

\bibitem{DB2}
\BIBentryALTinterwordspacing
(2023) Ibm db2. IBM. [Online]. Available: \url{https://www.ibm.com/products/db2}
\BIBentrySTDinterwordspacing

\bibitem{MariaDB}
\BIBentryALTinterwordspacing
(2023) Mariadb. MariaDB Foundation. [Online]. Available: \url{https://mariadb.org/}
\BIBentrySTDinterwordspacing

\bibitem{HybridClock2014}
S.~S. Kulkarni, M.~Demirbas, D.~Madappa, B.~Avva, and M.~Leone, ``Logical physical clocks,'' in \emph{Principles of Distributed Systems: 18th International Conference, OPODIS 2014, Cortina d’Ampezzo, Italy, December 16-19, 2014. Proceedings 18}.\hskip 1em plus 0.5em minus 0.4em\relax Springer, 2014, pp. 17--32.

\bibitem{HuaweiTaishan1}
\BIBentryALTinterwordspacing
(2023) Taishan 2480 server. Huawei. [Online]. Available: \url{https://e.huawei.com/en/products/computing/kunpeng/taishan/taishan-2480-v2}
\BIBentrySTDinterwordspacing

\bibitem{euleros}
M.~Zhou, X.~Hu, and W.~Xiong, ``openeuler: Advancing a hardware and software application ecosystem,'' \emph{IEEE Software}, vol.~39, no.~2, pp. 101--105, 2022.

\bibitem{TPC-C}
\BIBentryALTinterwordspacing
(2010) Tpc benchmark c revision 5.11. TPC. [Online]. Available: \url{https://www.tpc.org/TPC_Documents_Current_Versions/pdf/tpc-c_v5.11.0.pdf}
\BIBentrySTDinterwordspacing

\bibitem{Sysbench}
\BIBentryALTinterwordspacing
A.~Kopytov. (2013) Sysbench homepage. [Online]. Available: \url{https://github.com/akopytov/sysbench}
\BIBentrySTDinterwordspacing

\bibitem{LZ4}
\BIBentryALTinterwordspacing
(2011) Lz4 - extremely fast compression. Collet, Y. [Online]. Available: \url{https://github.com/lz4/lz4}
\BIBentrySTDinterwordspacing

\bibitem{TCPcongestioncontrol}
\BIBentryALTinterwordspacing
A.~Toonk. (2020) Tcp bbr - exploring tcp congestion control. Medium. [Online]. Available: \url{https://atoonk.medium.com/tcp-bbr-exploring-tcp-congestion-control-84c9c11dc3a9}
\BIBentrySTDinterwordspacing

\bibitem{nagle1984congestion}
J.~Nagle, ``Congestion control in ip/tcp internetworks,'' \emph{SIGCOMM Comput. Commun. Rev.}, vol.~14, no.~4, p. 11–17, oct 1984.

\bibitem{trafficcontrol}
B.~Hubert, T.~Graf, G.~Maxwell, R.~van Mook, M.~van Oosterhout, P.~Schroeder, J.~Spaans, and P.~Larroy, ``Linux advanced routing \& traffic control,'' in \emph{Proceedings of the Ottawa Linux Symposium}, sn.\hskip 1em plus 0.5em minus 0.4em\relax Ottawa, Ontario, Canada: Ottawa Linux Symposium, 2002, pp. 213--222.

\bibitem{anadiotis2020}
\BIBentryALTinterwordspacing
A.-C. Anadiotis, R.~Appuswamy, A.~Ailamaki, I.~Bronshtein, H.~Avni, D.~Dominguez-Sal, S.~Goikhman, and E.~Levy, ``A system design for elastically scaling transaction processing engines in virtualized servers,'' \emph{Proc. VLDB Endow.}, vol.~13, no.~12, p. 3085–3098, aug 2020. [Online]. Available: \url{https://doi.org/10.14778/3415478.3415536}
\BIBentrySTDinterwordspacing

\bibitem{avni2020}
\BIBentryALTinterwordspacing
H.~Avni, A.~Aliev, O.~Amor, A.~Avitzur, I.~Bronshtein, E.~Ginot, S.~Goikhman, E.~Levy, I.~Levy, F.~Lu, L.~Mishali, Y.~Mo, N.~Pachter, D.~Sivov, V.~Veeraraghavan, V.~Vexler, L.~Wang, and P.~Wang, ``Industrial-strength oltp using main memory and many cores,'' \emph{Proc. VLDB Endow.}, vol.~13, no.~12, p. 3099–3111, aug 2020. [Online]. Available: \url{https://doi.org/10.14778/3415478.3415537}
\BIBentrySTDinterwordspacing

\bibitem{opengauss2021}
\BIBentryALTinterwordspacing
G.~Li, X.~Zhou, J.~Sun, X.~Yu, Y.~Han, L.~Jin, W.~Li, T.~Wang, and S.~Li, ``Opengauss: An autonomous database system,'' \emph{Proc. VLDB Endow.}, vol.~14, no.~12, p. 3028–3042, jul 2021. [Online]. Available: \url{https://doi.org/10.14778/3476311.3476380}
\BIBentrySTDinterwordspacing

\bibitem{ROS4}
D.~B. Lomet, ``High speed on-line backup when using logical log operations,'' \emph{ACM SIGMOD Record}, vol.~29, no.~2, pp. 34--45, 2000.

\bibitem{daudjee2004}
K.~Daudjee and K.~Salem, ``Lazy database replication with ordering guarantees,'' in \emph{Proceedings of the 20th International Conference on Data Engineering}, ser. ICDE '04.\hskip 1em plus 0.5em minus 0.4em\relax USA: IEEE Computer Society, 2004, p. 424.

\bibitem{HuaweiTaishan2}
\BIBentryALTinterwordspacing
(2023) Taishan 5280 server. Huawei. [Online]. Available: \url{https://e.huawei.com/mx/products/servers/taishan-server/taishan-5280-v2}
\BIBentrySTDinterwordspacing

\bibitem{HuaweiopenGauss}
\BIBentryALTinterwordspacing
(2023) opengauss relational database system. Huawei. [Online]. Available: \url{https://gitee.com/opengauss/openGauss-server}
\BIBentrySTDinterwordspacing

\bibitem{ConfluxDB}
\BIBentryALTinterwordspacing
P.~Chairunnanda, K.~Daudjee, and M.~T. \"{O}zsu, ``Confluxdb: Multi-master replication for partitioned snapshot isolation databases,'' \emph{Proc. VLDB Endow.}, vol.~7, no.~11, p. 947–958, jul 2014. [Online]. Available: \url{https://doi.org/10.14778/2732967.2732970}
\BIBentrySTDinterwordspacing

\bibitem{10.1145/1323293.1294281}
\BIBentryALTinterwordspacing
G.~DeCandia, D.~Hastorun, M.~Jampani, G.~Kakulapati, A.~Lakshman, A.~Pilchin, S.~Sivasubramanian, P.~Vosshall, and W.~Vogels, ``Dynamo: Amazon's highly available key-value store,'' \emph{SIGOPS Oper. Syst. Rev.}, vol.~41, no.~6, p. 205–220, oct 2007. [Online]. Available: \url{https://doi.org/10.1145/1323293.1294281}
\BIBentrySTDinterwordspacing

\bibitem{carousel}
\BIBentryALTinterwordspacing
X.~Yan, L.~Yang, H.~Zhang, X.~C. Lin, B.~Wong, K.~Salem, and T.~Brecht, ``Carousel: Low-latency transaction processing for globally-distributed data,'' in \emph{Proceedings of the 2018 International Conference on Management of Data}, ser. SIGMOD '18.\hskip 1em plus 0.5em minus 0.4em\relax New York, NY, USA: Association for Computing Machinery, 2018, p. 231–243. [Online]. Available: \url{https://doi.org/10.1145/3183713.3196912}
\BIBentrySTDinterwordspacing

\bibitem{TAPIR}
\BIBentryALTinterwordspacing
I.~Zhang, N.~K. Sharma, A.~Szekeres, A.~Krishnamurthy, and D.~R.~K. Ports, ``Building consistent transactions with inconsistent replication,'' \emph{ACM Trans. Comput. Syst.}, vol.~35, no.~4, dec 2018. [Online]. Available: \url{https://doi.org/10.1145/3269981}
\BIBentrySTDinterwordspacing

\bibitem{ren2019slog}
\BIBentryALTinterwordspacing
K.~Ren, D.~Li, and D.~J. Abadi, ``Slog: Serializable, low-latency, geo-replicated transactions,'' \emph{Proc. VLDB Endow.}, vol.~12, no.~11, p. 1747–1761, jul 2019. [Online]. Available: \url{https://doi.org/10.14778/3342263.3342647}
\BIBentrySTDinterwordspacing

\bibitem{STAR2019}
\BIBentryALTinterwordspacing
Y.~Lu, X.~Yu, and S.~Madden, ``{STAR},'' \emph{Proceedings of the {VLDB} Endowment}, vol.~12, no.~11, pp. 1316--1329, jul 2019. [Online]. Available: \url{https://doi.org/10.14778%2F3342263.3342270}
\BIBentrySTDinterwordspacing

\bibitem{Huygens2018}
\BIBentryALTinterwordspacing
Y.~Geng, S.~Liu, Z.~Yin, A.~Naik, B.~Prabhakar, M.~Rosenblum, and A.~Vahdat, ``Exploiting a natural network effect for scalable, fine-grained clock synchronization,'' in \emph{15th USENIX Symposium on Networked Systems Design and Implementation (NSDI 18)}.\hskip 1em plus 0.5em minus 0.4em\relax Renton, WA: USENIX Association, Apr. 2018, pp. 81--94. [Online]. Available: \url{https://www.usenix.org/conference/nsdi18/presentation/geng}
\BIBentrySTDinterwordspacing

\bibitem{BBR}
N.~Cardwell, Y.~Cheng, S.~H. Yeganeh, and V.~Jacobson, ``Bbr congestion control,'' \emph{IETF Draft draft-cardwell-iccrg-bbr-congestion-control-00}, 2017.

\bibitem{NAGEL}
G.~Minshall, Y.~Saito, J.~C. Mogul, and B.~Verghese, ``Application performance pitfalls and tcp's nagle algorithm,'' \emph{ACM SIGMETRICS Performance Evaluation Review}, vol.~27, no.~4, pp. 36--44, 2000.

\bibitem{paxos}
L.~Lamport, ``Paxos made simple,'' \emph{ACM SIGACT News (Distributed Computing Column) 32, 4 (Whole Number 121, December 2001)}, pp. 51--58, 2001.

\end{thebibliography}
\nocite{*}

\end{document}